\documentclass[aps,prb,reprint,groupedaddress]{revtex4-1}

\usepackage{amssymb}
\usepackage{amsmath}
\usepackage{graphicx}
\usepackage{epsfig}
\usepackage{subfigure}

\begin{document}

\title{Peak effect in a superconductor/normal metal strip being in vortex-free state}

\author{P. M. Marychev}
\email[Corresponding author: ]{marychevpm@ipmras.ru}
\author{D. Yu. Vodolazov}
\affiliation{Institute for Physics of Microstructures, Russian
Academy of Sciences, Nizhny Novgorod, 603950 Russia}

\date{\today}

\begin{abstract}
We theoretically predict that the critical current $I_c$ and
magnetization $M$ of hybrid superconductor/normal-metal (SN) strip
may have nonmonotonous dependence on perpendicular magnetic field
- so called peak effect. In contrast to familiar peak effect,
which is connected either with vortex entry to the superconductor
or with peculiarities of vortex pinning, the found phenomenon
exists at low fields, in the vortex-free (Meissner) phase. We
argue that the effect appears at specific parameters of studied
hybrid structure when its in-plane current-supervelocity relation
has two maxima. We expect that the same peak effect may exist in
two-band superconductors (like MgB$_2$) where similar
current-supervelocity dependence was predicted at low
temperatures.
\end{abstract}

\maketitle

The influence of a perpendicular magnetic field $H$ on transport
properties of type-II superconductors has been the subject of
numerous studies. Usually the critical current $I_c$ of bulk
superconductors is determined mainly by pinning of vortices on
defects and it monotonically decreases with increasing $H$.
However in conventional low-$T_c$ superconductors there has been
observed a peak in $I_c(H)$ just below the upper critical field
$H_{c2}$ (see for example
[\onlinecite{DeSorbo-1964,Isino-1988,Kokubo-2007}]). The peak in
$I_c(H)$ is also accompanied by a peak in the dependence of
magnetization $M$ on $H$ and this phenomenon is called as the peak
effect. The peak effect near $H_{c2}$ is explained by a softening
of the vortex lattice [\onlinecite{Pippard-1967,Larkin-1979}].
Also there was discovered the peak located significantly below
$H_{c2}$ both in the low-$T_c$
[\onlinecite{Banerjee-2000,Lortz-2007}] and high-$T_c$
superconductors [\onlinecite{Zhukov-1995,Chen-2011}]. The origin
of this type of peak is explained by the transition from a
quasiordered vortex lattice to an amorphous vortex glass state.

In a homogeneous superconducting strip the critical current could
be determined not by the bulk pinning of vortices but by the edge
barrier for their entrance
[\onlinecite{Kupriyanov_1974,Benkraouda_1998,Maksimova-2001,Plourde-2001}].
Usually effect of edge barrier is pronounced in a thin
strip/bridge with thickness $d_S$ less than the London magnetic field
penetration depth $\lambda$ and at relatively low magnetic fields
when there is no dense vortex lattice
[\onlinecite{Andratskii-1974,Fuchs-1998,Plourde-2001,Ilin-2014}].
In relatively narrow strip (with width $W \ll
\Lambda=\lambda^2/d_S$) one may observe peak in $I_c(H)$ near the
field for first vortex entry
[\onlinecite{Eisenmenger-2004,Lin-2013,Ilin-2014}] (the same peak
has been observed in thin Pb/In and Nb strips placed in parallel
magnetic field [\onlinecite{Yamashita_1976,Ichkitidze_1981}]). It
originates from the entrance of the vortex row at some field which
does not exit the strip and it prevents subsequent vortex entry
[\onlinecite{Shmidt_1970}]. It is interesting that competition of
the bulk pinning and edge barrier also may lead, at some
parameters, to the peak effect at low magnetic fields, as it was
predicted in Ref. [\onlinecite{Elistratov-2002}].

Here we argue that peak in $I_c(H)$ and $M(H)$ may arise even in
the vortex-free state. Below we show that it could be realized,
for example, in a hybrid superconductor/normal metal (SN) thin
strip with large ratio of resistivities of S and N layers
$\rho_S/\rho_N \gg 1$ in the normal state. In Ref.
[\onlinecite{Vodolazov-2018}] it has been shown that dependence of
superconducting sheet current density $J_s$ (in ordinary S strip
$J_s=j_sd_S$ where $j_s$ is a superconducting current density) on
supervelocity $v_s$ or supermomentum $\hbar q=\hbar (\nabla\varphi
+2\pi A/\Phi_0)\sim v_s$ (here $\varphi$ is the phase of
superconducting order parameter and $A$ is the vector potential)
may have two maxima at low temperature. The first maximum at small
$q$ is connected with suppression of the proximity induced
superconductivity in the N layer, while the second maximum at
large $q$ comes from suppression of superconductivity in the S
layer. The predicted dependence is rather different from $J_s(q)$
of ordinary one-band superconductor, which has only one maximum,
but it resembles the dependence $J_s(q)$ for two-band
superconductors [\onlinecite{Koshelev,Nicol}]. In that case
different maxima correspond to destruction of superconductivity in
different bands.

\begin{figure}[hbt]
\includegraphics[width=0.9\linewidth]{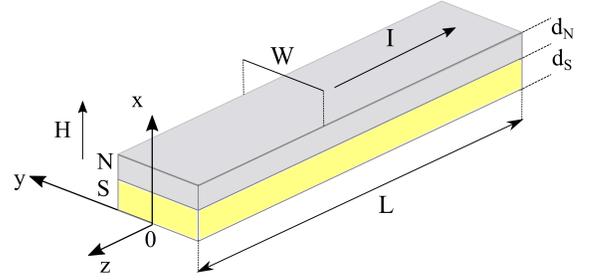}
\caption{\label{Fig:Sys} Sketch of SN strip with transport current
$I$ and placed in perpendicular magnetic field $H$.}
\end{figure}

Our model system is shown in Fig.~\ref{Fig:Sys}. The SN strip with
width $W$ has two layers: superconducting one with thickness $d_S$
and the normal metal layer with thickness $d_N$. In calculations
we use one and two-dimensional Usadel equation for normal
$g=cos\Theta$ and anomalous $f=sin\Theta\exp(i\varphi)$
quasi-classical Green functions, assuming that angle $\Theta$
depends only on x and y and length of the SN strip $L\to\infty$
(equations and details of the model are presented in Appendix and could be found in Ref.
[\onlinecite{Vodolazov-2018}]). Our model is not able to take into
account vortex states so we consider here only the Meissner
(vortex-free) state. We consider narrow strip with width smaller
than the magnetic field penetration depth $\Lambda$ of the single
S layer to neglect the contribution of screening currents to
vector potential which we choose as: ${\bf A}=(0,0,Hy)$. In our
model we assume that current reaches the critical value when
$q(y=W/2)=q_c$, where $q(y)=\nabla\varphi +2\pi A(y)/\Phi_0$
($\nabla\varphi(y)=const$) and $q_c$ is the critical value of $q$
corresponding to the reaching depairing current density at the
edge. This condition corresponds to instability of the Meissner
state with respect to vortex entry [\onlinecite{Vodolazov_2003}].

To find $I_c(H)$ we numerically solve either 1D or 2D Usadel
equations (see Appendix). In 1D model we split SN strip to
filaments with width $\xi_c$ and assume that $J_s(y)=\int
j_s(x,y)dx=\int j_s(x,q(y))dx=J_s(q(y))$ ($q$ depends on
y-coordinate of filament) and may be found from solution of 1D
Usadel equation (in this case $\Theta$ has dependence only on x
coordinate). Then we calculate $I_c=\int J_s(q(y))dy$. In 2D model
we solve 2D Usadel equation with given $q(y)$ and find $I_c=\int
j_s(x,y)dxdy$ ($\Theta$ depends both on x and y). The difference
between these approaches is that in 1D model we neglect proximity
effect between adjacent filaments which brings the difference
between $J_s(y)$ and $J_s(q(y))$. We expect that the filament
model gives quantitatively correct results when $W\gg
\xi_N=\sqrt{\hbar D_N/k_BT}$ [\onlinecite{Ustavschikov-2021}],
where $D_N$ is a diffusion coefficient in N layer.

In calculations we normalize lengths in units of
$\xi_c=\sqrt{\hbar D_S/k_BT_{c0}}$, where $T_{c0}$ is the critical
temperature and $D_S$ is the diffusion coefficient of S layer.
Sheet current density $J_s$ is normalized in units of depairing
sheet current density $J_{dep}(0)=I_{dep}(0)/d_S$ of S layer at
$T=0$ and the magnetic field is measured in units of
$H_s=\Phi_0/2\pi W\xi_c$ (this field is about of first vortex
entry field [\onlinecite{Maksimova-2001,Plourde-2001}] to the strip at $I=0$). We choose ratio of
resistivities (diffusion coefficients) $\rho_S/\rho_N=D_N/D_S=100$
which corresponds to NbN, NbTiN, MoN or MoSi as a superconductor
and Ag, Cu or Au as a normal metal.

\begin{figure}[]
\includegraphics[width=\linewidth]{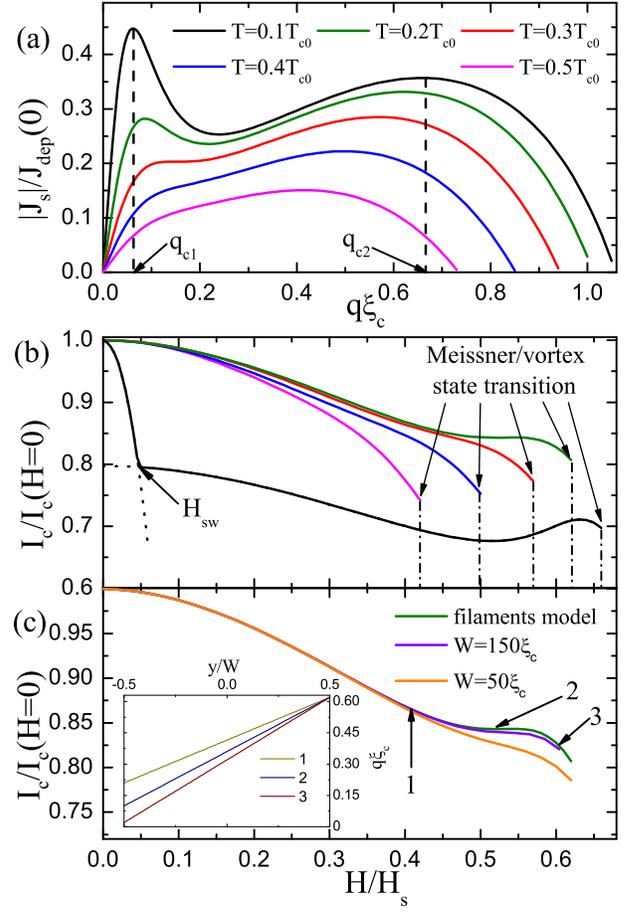}
 \caption{\label{Fig:i(h)-sn}(a) Dependence of the amplitude of sheet superconducting current density $J_s$ on
$q$ in SN strip calculated at different temperatures $T$ in 1D
Usadel model. (b) Calculated $I_c(H)$ of SN strip at different $T$
(filaments model). (c) Calculated $I_c(H)$ in filaments and 2D
Usadel models at $T=0.2T_{c0}$. In the inset we show the spatial
distribution of $q(y)$ over SN strip at different fields marked by
numbers 1--3. SN strip has the following parameters: $d_S=2\xi_c$,
$d_N=4\xi_c$ and $\rho_S/\rho_N=100$.}
\end{figure}

In Fig.~\ref{Fig:i(h)-sn}(a) we show temperature evolution of
$|J_s|(q)$ (it was found from solution of 1D Usadel equation)
which is used for calculation of the critical current in the
filaments model. With decreasing temperature the dependence
$|J_s|(q)$ transforms from the ordinary one (with one maximum) to
the dependence with two maxima located at $q=q_{c1}$ and
$q=q_{c2}$ (note qualitative similarity with $|J_s|(q)$ for the
two-band superconductor MgB$_2$ [\onlinecite{Koshelev,Nicol}]).
The first maximum comes from the suppression of proximity-induced
superconductivity in the N layer where
$q_c=q_{c1}\propto\sqrt{1/D_N}$. The second maximum comes from the
suppression of superconductivity in the S layer where $q_c=
q_{c2}\propto\sqrt{1/D_S}\gg q_{c1}$. The increase of the
amplitude of first maximum at low temperatures is explained by the
enhancement of the proximity-induced superconductivity while the
'strength' of the intrinsic superconductivity in S layer is
already saturated and amplitude of second maximum weakly depends
on temperature at low T. As it is discussed in Ref.
[\onlinecite{Vodolazov-2018}] such an evolution in $J_s(q)$ should
lead to the kink on dependence $I_c(T)$ at low T (when
$|J_s|(q_{c1})$ becomes larger than $|J_s|(q_{c2})$). The same
kink is also predicted in Refs. [\onlinecite{Koshelev,Nicol}] for
MgB$_2$ and it is caused by the similar change of $J_s(q)$ with
temperature.

The dependence $I_c(H)$ (see Fig.~\ref{Fig:i(h)-sn}(b)) changes
with the temperature accordingly to transformation of $J_s(q)$.
Indeed, external magnetic field changes the distribution of $q$
across the strip (see inset in Fig.~\ref{Fig:i(h)-sn}(c)). When the
width of the SN strip is much larger than $\xi_N$ one may assume
that local $J_s(y)$ is determined only by local $q(y)$. With
increasing magnetic field $q$ decreases in the strip (except at
the edge $y=W/2$) and it leads to monotonous decrease of $|J_s|$ and
critical current $I_c=\int J_s dy$ when dependence $|J_s|(q)$ has
only one maximum. However, with decreasing temperature the
additional maximum appears at low $q$. At first it leads to
flattening of $I_c(H)$ (see Fig.~\ref{Fig:i(h)-sn}(b) at $T=0.2
T_{c0}$) because of flattening of $J_s(q)$. When the height of the
first maximum becomes larger than the second one the dependence
$I_c(H)$ changes drastically (see Fig.~\ref{Fig:i(h)-sn}(b) at
T=0.1 T$_{c0}$). At low fields $I_c$ drops fast with increase of
$H$ because of much smaller value of $q(W/2)=q_c=q_{c1} \ll
q_{c2}$. At some field (marked by $H_{sw}$ in
Fig.~\ref{Fig:i(h)-sn}(b)) the SN strip transit to the state with
larger $q(W/2)=q_c=q_{c2}$ because in this case SN strip may carry
larger critical current [\onlinecite{selfref}]. At $H>H_{sw}$ the
peak in $I_c(H)$ appears which is consequence of the first maximum
in $|J_s|(q)$.

At some magnetic field $q$ and $J_s(q)$ change the sign at
$y=-W/2$. It means that vortices, which enter at opposite edge
($y=W/2$) cannot exit the SN strip. In ordinary S strip it leads
to the peak in $I_c(H)$ [\onlinecite{Shmidt_1970,Vodolazov-2013}].
We expect similar behavior in SN strip too. Because vortex states
cannot be described by the used model we are bounded by the field
$q_{c2}\Phi_0/2\pi W$ at which $q(-W/2)=0$.

Discussed above features could be seen only for sufficiently wide
strips. In a relatively narrow strip with $W\lesssim \xi_N$ the
proximity effect from the adjacent regions plays important role
and nonmonotonous behavior is smeared out (see
Fig.~\ref{Fig:i(h)-sn}(c)).

\begin{figure}[]
\includegraphics[width=1.07\linewidth]{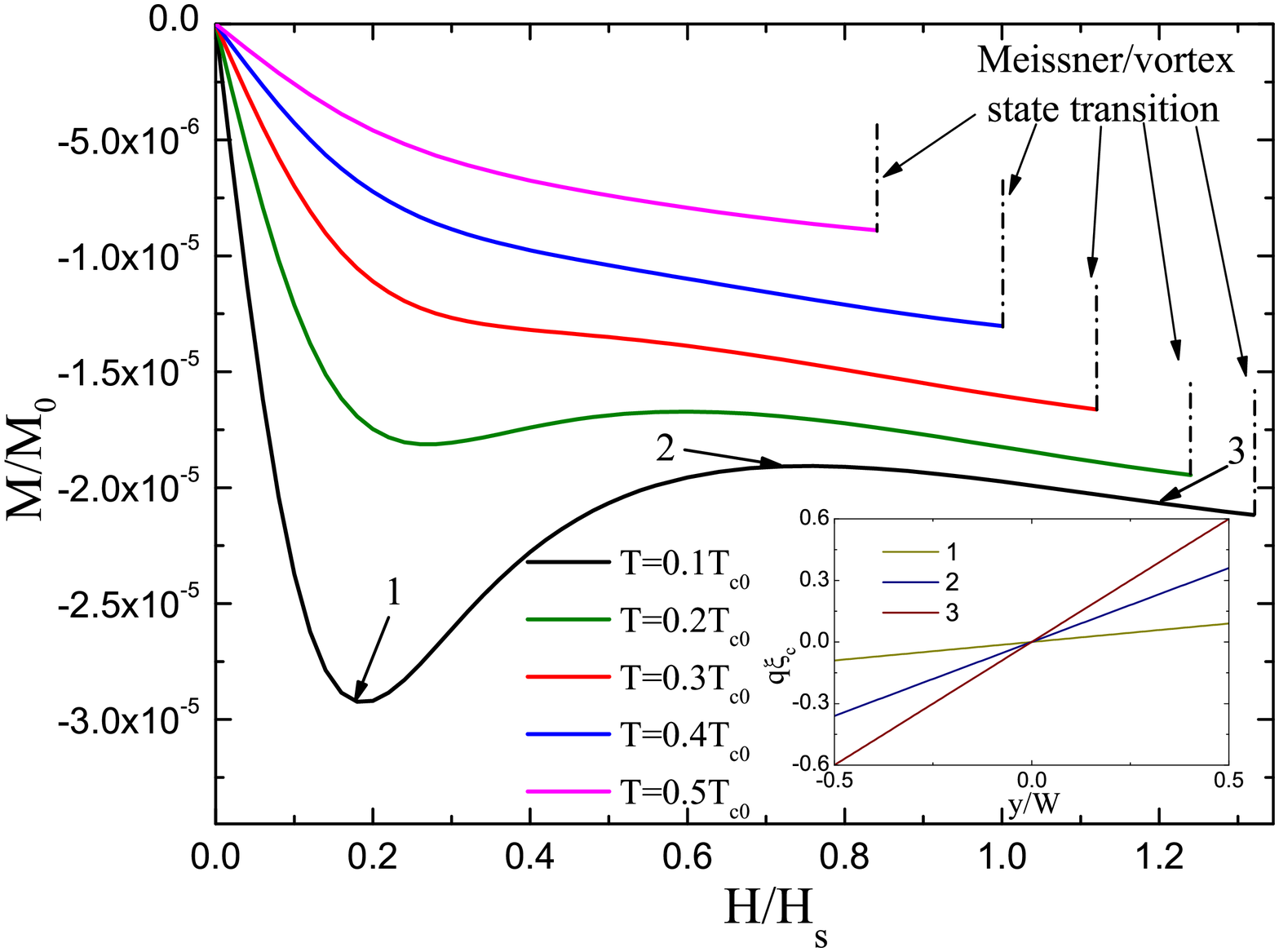}
 \caption{\label{Fig:m(h)-sn} The magnetization curves of the SN strip
calculated in 2D model at different temperatures. Magnetization
$M$ is measured in units of $M_0=\Phi_0/2\pi\xi_c^2$. In the inset
we show the distribution $q(y)$ over SN strip at $T=0.1T_{c0}$ and
different fields marked by numbers 1--3. SN strip has the width
$W=80\xi_c$ and the other parameters are as in
Fig.~\ref{Fig:i(h)-sn}.}
\end{figure}

It is known, that in the ordinary superconductors the peak in
$I_c(H)$ is followed by the peak in the magnetization curve $M(H)$
(or vise versus). In SN strip we also find the peak in
$M(H)=\int\int [{\bf r\times j_s}] dxdy/(2c(d_S+d_N)W)$ (see
Fig.~\ref{Fig:m(h)-sn}). But in contrast to the ordinary peak
effect the peaks are located at different fields for $I_c(H)$ and
$M(H)$ dependencies. The reason for this is following. In absence
of the current the evolution of $q(y)$ with increasing of $H$ is
different to the situation with current $I=I_c(H)$ (see inset in
Fig.~\ref{Fig:m(h)-sn} - in this case $q(y)=2\pi A(y)/\Phi_0$). It
results to larger screening currents at low fields than at high
$H$ (at $T=0.1T_{c0}$) and peak is located at lower field. Here we
stop calculations at the magnetic field $H=\Phi_0 q_{c2}/(\pi W)$
when $|q(\pm W/2)|=q_{c2}$ and we expect vortex entrance to the SN
strip.

We believe that the same effect should exist in SS' bilayer where
S' is a superconductor which has large diffusion coefficient (low
resistivity in the normal state, for example Al, Pb or Sn). Due to
large $D_{S'}$ the superconductivity in S' layer should be
destroyed at smaller $q$ and the current-supervelocity dependence
will have two maxima at proper choice of $d_S$, $d_{S'}$ and
temperature. Because qualitatively similar $J_s(q)$ dependence was
predicted for two-band superconductor MgB$_2$ (see
[\onlinecite{Koshelev,Nicol}]), and, hence, a peak or plateau in
$I_c(H)$ should be observed in MgB$_2$ thin strip at fields
$\lesssim q_{c2}\Phi_0/2\pi W$. But important condition for
experimental observation of predicted effect is approaching of
$I_c(H=0)$ to the depairing current of SN, SS' or MgB$_2$. Most
easily critical current about of $I_{dep}$ could be probably reached
in SN system how it has been demonstrated recently for MoN/Cu
strip [\onlinecite{Ustavschikov-2021}]. However, in MgB$_2$
strips/bridges depairing current has not been reached yet. In Refs.
[\onlinecite{Kunchur-2003,Zhuang-2008,Novoselov-2017}] $I_c\simeq
15-30\%$ of the depairing current was claimed which is not large
enough for observation of the predicted peak effect.

To conclude, we hope that the experimental observation of the peak
or plateau on $I_c(H)$ and/or $M(H)$ dependencies at low fields
would indirectly confirm existence of two peaks in $J_s(q)$
dependence in SN, SS' hybrid structures or many-band
superconducting materials.

\begin{acknowledgments}
We thank A. Yu. Aladyshkin for helpful discussion.
\end{acknowledgments}

\appendix*
\section{Usadel model}

To calculate superconducting properties of the SN strip, we use
the Usadel model for normal $g=cos\Theta$ and anomalous
$f=sin\Theta\exp(i\varphi)$ quasi-classical Green functions inside
both S and N layers. We neglect the dependence of $\Theta$ on the
longitudinal coordinate z since length of the SN strip
$L\to\infty$ and the system is uniform in this direction.
Therefore we use the one-dimensional (1D) Usadel equation

\begin{equation}
 \label{usadel-1d}
 \frac{\hbar D}{2}\frac{\partial^2\Theta}{\partial
x^2}-\left(\hbar\omega_n+\frac{\hbar D}{2}q^2\cos
\Theta\right)\sin\Theta+\Delta\cos\Theta=0
\end{equation}
and the two-dimensional (2D) Usadel equation
\begin{equation}
 \label{usadel-2d}
 \begin{split}
 \frac{\hbar D}{2}\left(\frac{\partial^2\Theta}{\partial
x^2}+\frac{\partial^2\Theta}{\partial
y^2}\right)-\left(\hbar\omega_n+\frac{\hbar D}{2}q^2\cos
\Theta\right)\sin\Theta \\
+\Delta\cos\Theta=0.
\end{split}
\end{equation}
Here $D$ is a diffusion coefficient ($D=D_S$ and $D=D_N$ in superconducting and normal layers, respectively), $\hbar \omega_n = \pi k_BT(2n+1)$ are the
Matsubara frequencies ($n$ is an integer number), $\Delta$ is the superconducting order parameter, which is nonzero only in the S layer. Coordinate axes are presented in Fig.~\ref{Fig:Sys}. $\Delta$ should satisfy the
self-consistency equation

\begin{equation}
\label{self-cons}
\Delta \ln\left(\frac{T}{T_{c0}}\right)=2\pi k_B
T\sum_{\omega_n
>0} \left(\sin\Theta_S - \frac{\Delta}{\hbar\omega_n}\right),
\end{equation}
where $T_{c0}$ is the critical temperature of single S layer
in the absence of magnetic field. Equations (\ref{usadel-1d},\ref{usadel-2d})
is supplemented by the Kupriyanov-Lukichev boundary conditions
between layers [\onlinecite{JETP-1988}] with fully transparent
interfaces
  \begin{equation}
  \nonumber
    \label{boundary}
    \left.D_S\frac{d\Theta_S}{dx}\right|_{x=d_S-0}=\left.D_N\frac{d\Theta_N}{dx}\right|_{x=d_S+0}.
    \end{equation}
On the interfaces between the system and vacuum we use $d\Theta/dn=0$.

The superconducting current density is calculated as
\begin{equation}
 \label{current}
 j_s(x,y)=\frac{2\pi k_BT}{e\rho}q\sum_{\omega_n > 0}\sin^2\Theta,
\end{equation}
where $\rho$ is the resistivity of corresponding layer. To find $j_s(x,y)$, we numerically solve either equation
(\ref{usadel-1d}) or (\ref{usadel-2d}) and equation (\ref{self-cons}). Equations are solved by an iteration procedure using the Newton method combined with
a tridiagonal matrix algorithm. Obtained $\Theta (x,y)$ is inserted in equation (\ref{self-cons}) to find $\Delta$ and then
iterations repeat until the self-consistency is achieved.


\begin{thebibliography}{35}
\bibitem{selfref} Because of jump-like change of $q$ this transition is of the
first kind and there is a kink on dependence $I_c(H)$. During the
transition the electric field should appear which provides change
of $q$ everywhere in the SN strip. How this switching happens is
an interesting question but it is beyond of scope of our work
because it needs consideration of the time-dependent problem.
\end{thebibliography}
\end{document}